\documentclass[11pt]{article}
\usepackage{amsmath,amsthm,latexsym,amssymb,amsfonts,epsfig,psfrag}
\usepackage{color}

\makeatletter \@addtoreset{equation}{section} \makeatother

\def\be{\begin{equation}}
\def\ee{\end{equation}}
\def\ba{\begin{eqnarray}}
\def\ea{\end{eqnarray}}

\newcommand\nn{\nonumber}
\newcommand\q{\quad}
\def\Nl{{\mathchoice
{\setbox0=\hbox{$\displaystyle\rm N$}\hbox{\hbox to0pt
{\kern0.4\wd0\vrule height0.9\ht0\hss}\box0}}
{\setbox0=\hbox{$\textstyle\rm N$}\hbox{\hbox to0pt
{\kern0.4\wd0\vrule height0.9\ht0\hss}\box0}}
{\setbox0=\hbox{$\scriptstyle\rm N$}\hbox{\hbox to0pt
{\kern0.4\wd0\vrule height0.9\ht0\hss}\box0}}
{\setbox0=\hbox{$\scriptscriptstyle\rm N$}\hbox{\hbox to0pt
{\kern0.4\wd0\vrule height0.9\ht0\hss}\box0}}}}
\def\Zl{{\mathchoice
{\setbox0=\hbox{$\displaystyle\rm Z$}\hbox{\hbox to0pt
{\kern0.4\wd0\vrule height0.9\ht0\hss}\box0}}
{\setbox0=\hbox{$\textstyle\rm Z$}\hbox{\hbox to0pt
{\kern0.4\wd0\vrule height0.9\ht0\hss}\box0}}
{\setbox0=\hbox{$\scriptstyle\rm Z$}\hbox{\hbox to0pt
{\kern0.4\wd0\vrule height0.9\ht0\hss}\box0}}
{\setbox0=\hbox{$\scriptscriptstyle\rm Z$}\hbox{\hbox to0pt
{\kern0.4\wd0\vrule height0.9\ht0\hss}\box0}}}}
\def\Ql{{\mathchoice
{\setbox0=\hbox{$\displaystyle\rm Q$}\hbox{\hbox to0pt
{\kern0.4\wd0\vrule height0.9\ht0\hss}\box0}}
{\setbox0=\hbox{$\textstyle\rm Q$}\hbox{\hbox to0pt
{\kern0.4\wd0\vrule height0.9\ht0\hss}\box0}}
{\setbox0=\hbox{$\scriptstyle\rm Q$}\hbox{\hbox to0pt
{\kern0.4\wd0\vrule height0.9\ht0\hss}\box0}}
{\setbox0=\hbox{$\scriptscriptstyle\rm Q$}\hbox{\hbox to0pt
{\kern0.4\wd0\vrule height0.9\ht0\hss}\box0}}}}
\def\Rl{{\mathchoice
{\setbox0=\hbox{$\displaystyle\rm R$}\hbox{\hbox to0pt
{\kern0.4\wd0\vrule height0.9\ht0\hss}\box0}}
{\setbox0=\hbox{$\textstyle\rm R$}\hbox{\hbox to0pt
{\kern0.4\wd0\vrule height0.9\ht0\hss}\box0}}
{\setbox0=\hbox{$\scriptstyle\rm R$}\hbox{\hbox to0pt
{\kern0.4\wd0\vrule height0.9\ht0\hss}\box0}}
{\setbox0=\hbox{$\scriptscriptstyle\rm R$}\hbox{\hbox to0pt
{\kern0.4\wd0\vrule height0.9\ht0\hss}\box0}}}}
\def\Cl{{\mathchoice
{\setbox0=\hbox{$\displaystyle\rm C$}\hbox{\hbox to0pt
{\kern0.4\wd0\vrule height0.9\ht0\hss}\box0}}
{\setbox0=\hbox{$\textstyle\rm C$}\hbox{\hbox to0pt
{\kern0.4\wd0\vrule height0.9\ht0\hss}\box0}}
{\setbox0=\hbox{$\scriptstyle\rm C$}\hbox{\hbox to0pt
{\kern0.4\wd0\vrule height0.9\ht0\hss}\box0}}
{\setbox0=\hbox{$\scriptscriptstyle\rm C$}\hbox{\hbox to0pt
{\kern0.4\wd0\vrule height0.9\ht0\hss}\box0}}}}
\def\Hl{{\mathchoice
{\setbox0=\hbox{$\displaystyle\rm H$}\hbox{\hbox to0pt
{\kern0.4\wd0\vrule height0.9\ht0\hss}\box0}}
{\setbox0=\hbox{$\textstyle\rm H$}\hbox{\hbox to0pt
{\kern0.4\wd0\vrule height0.9\ht0\hss}\box0}}
{\setbox0=\hbox{$\scriptstyle\rm H$}\hbox{\hbox to0pt
{\kern0.4\wd0\vrule height0.9\ht0\hss}\box0}}
{\setbox0=\hbox{$\scriptscriptstyle\rm H$}\hbox{\hbox to0pt
{\kern0.4\wd0\vrule height0.9\ht0\hss}\box0}}}}
\def\Ol{{\mathchoice
{\setbox0=\hbox{$\displaystyle\rm O$}\hbox{\hbox to0pt
{\kern0.4\wd0\vrule height0.9\ht0\hss}\box0}}
{\setbox0=\hbox{$\textstyle\rm O$}\hbox{\hbox to0pt
{\kern0.4\wd0\vrule height0.9\ht0\hss}\box0}}
{\setbox0=\hbox{$\scriptstyle\rm O$}\hbox{\hbox to0pt
{\kern0.4\wd0\vrule height0.9\ht0\hss}\box0}}
{\setbox0=\hbox{$\scriptscriptstyle\rm O$}\hbox{\hbox to0pt
{\kern0.4\wd0\vrule height0.9\ht0\hss}\box0}}}}
\newcommand{\bd}{\mathbf d}

\title{From the discrete to the continuous -- \\ towards a cylindrically consistent dynamics}
\author{ Bianca Dittrich\\
\small Perimeter Institute for Theoretical Physics,\\
\small 31 Caroline St. N, Waterloo, ON N2L 2Y5, Canada\\
\small and\\
\small   MPI for Gravitational Physics,\\
 \small Am M\"uhlenberg 1, D-14476 Potsdam, Germany \\
}

\date{}

\begin{document}

\maketitle

\begin{abstract}
Discrete models usually represent approximations to continuum physics. Cylindrical consistency provides a framework in which discretizations mirror exactly the continuum limit. Being a standard tool for the kinematics of loop quantum gravity we propose a coarse graining procedure that aims at constructing a cylindrically consistent dynamics in the form of transition amplitudes and Hamilton's principal functions.  The coarse graining procedure, which is motivated by tensor network renormalization methods, provides a systematic approximation scheme towards this end.

A crucial role in this coarse graining scheme is played by embedding maps that allow the interpretation of discrete boundary data as continuum configurations. These embedding maps should be selected according to the dynamics of the system, as a choice of embedding maps will determine a truncation of the renormalization flow. 
\end{abstract}


\section{Introduction}

Discretizations  have become a leading tool for the construction of  quantum gravity models \cite{lollreview}, for instance in loop quantum gravity \cite{thomasbook,bahrreview}, spin foams \cite{carlobook,perezreview} or Regge gravity \cite{regge}. Indeed discrete models provide a very effective method to access non-perturbative physics. 


Given a discretization, the question arises how to relate observables from the discrete model to observables in the corresponding continuum limit (should it exist). Usually this question is handled by considering a family of discretizations, in which we can consider the refinement limit. Observables computed from a given discretizations will in general be approximations to corresponding observables in the continuum limit. 

An alternative view is provided by the concept of cylindrical consistency, which is a key technique in the construction of the continuum Hilbert space of loop quantum gravity \cite{al}. The idea here is that discretizations are not seen as approximation but as a selection of a certain set of degrees of freedom, for which the discrete model should give the same predictions as the continuum model. More precisely discretizations come with a (partial) ordering into finer and coarser and embedding maps\footnote{More precisely the configuration spaces or Hilbert spaces associated to coarser discretizations are embedded into those for finer discretizations.} of coarser into finer discretizations. Cylindrical consistency demands that observables which can be represented in a given discretization should not depend on the choice of finer discretization into which this given discretization is embedded. In other words,  a prediction drawn from a given discretization should be also valid for any refined discretization or the continuum limit.

So far this concept has been successfully applied to the construction of the kinematical Hilbert space \cite{al,thomasbook}, including observables and the Hamiltonian constraints \cite{qsd}. The question arises whether we can implement this concept for the dynamics, for instance in the form of a path integral (measure) as suggested in \cite{benloops} see also \cite{zapata} for related ideas. This would be equivalent to a cylindrical consistent construction of transition amplitudes (or the physical Hilbert space). On the classical level we would ask for cylindrically consistent family of Hamilton's principal functions which are the classical analogues of transition amplitudes  \cite{carloham}. 

In this way cylindrical consistency not only provides the continuum limit of a given theory, it also allows via the embedding maps a precise interpretation of discrete configurations (states) as elements of the continuum configuration (Hilbert) space. 

The concept of a cylindrically consistent dynamics is actually very similar to the concept of perfect actions \cite{hasenfratz} or perfect discretizations. These were suggested for gravitational systems in \cite{alg,bahrdittrich2} and constructed for a number of examples in \cite{bahrdittrich2,song,seb1,seb2} to address the problem of broken diffeomorphism symmetry in generic discretizations of gravity \cite{dittrichreview,bahrdittrich1}.

Indeed recent work \cite{bahrdittrich2,seb1,ditt, zako} has shown that demanding diffeomorphism symmetry is a very strong requirement: there are many examples where it is equivalent to demanding triangulation (or more generally discretization) independence.\footnote{This notion of diffeomorphism invariance implies that the positions of vertices in a given solution are gauge parameters and can be chosen freely. But if we are allowed to move vertices we can produce very densely triangulated regions and very coarse triangulated regions. In the extreme case one can move vertices on top of each other thus even reducing the number of vertices in a given discretization, or the number of simplices in a given triangulation.}  This includes coarse graining the triangulation or making one region much denser triangulated than another region. Such a notion implies that the discrete theory should correctly reproduce physics on all scales, not only on scales large compared to the discretization scale \cite{zako}. In other words such a discretization has to mirror continuum theory exactly, which is the defining property of a perfect discretization.

A way to construct such perfect discretizations is as fixed points of a (real space) Wilsonian renormalization flow. Such a flow generates non-local couplings (in non-topological theories), see for instance \cite{bietenholz,song}. These non-local couplings are crucial in order to obtain a non--topological theory which is triangulation independent.

The non--local couplings make the theory also computationally very challenging. As one basically attempts to solve the dynamics of the theory, a systematic framework that would improve the discretization in steps, and also allows a control on the amount of  non-local couplings, is advisable. In this work we will propose such a framework. As a key input we introduce boundaries, the fineness of the boundary data will determine the quality of the approximation. This already indicates that we will have a partial ordering into finer and coarser boundaries and corresponding embedding maps. This will allow us to formulate the aim of achieving a perfect discretization as finding a cylindrically consistent effective action or Hamilton's principal function/ path integral measure. As we will see the choice of embedding maps will be essential and should be determined by the dynamics of the system.

Another view point of how this proposal differs from the usual understanding of perfect discretizations, is that in the latter we keep the building blocks of the discretization simple but allow for non--local couplings between these building blocks. Here we will introduce more and more complicated building blocks with finer and finer boundaries. The couplings between these building blocks will still be local, i.e. only variables associated to neighbouring building blocks will be coupled. Note however that the number of such variables will be growing and that the couplings in-between the variables associated to one building block will be non--local (if we see this building block as an extended object).  In fact this shift of non--localities from in-between building blocks to introducing more and more variables to one building block is inspired by tensor network renormalization \cite{levin,wen,eckert}. There the path integral amplitude (together with the measure) associated to a certain region or building block is encoded  into a tensor. Summing over (bulk) variables correspond to the contraction of tensors into a new tensor, which typically is of higher rank, representing more complicated boundary data. The key question is then how to approximate a tensor of higher rank with a tensor of the original rank, that is one has to choose a truncation of the renormalization flow.  Here we will put forward the question whether and how we could use the embedding maps, which underly cylindrical consistency, for such a truncation. This relation of embedding maps to truncations of the renormalization flow makes it obvious that the embedding maps should be determined according to the dynamics of the system.

~\\
In the next sections \ref{scalar} and \ref{pot} we will introduce the concept with the help of an example, the scalar field with and without potential in two dimensions. In the following sections \ref{larger} and \ref{other} we will consider variations of the coarse graining scheme and compare two different choices for the embedding maps. The coarse graining scheme will be formalized in section \ref{cyl}, which will also shortly review the concept of cylindrical consistency. In section \ref{quantum} we will extend the procedure to the quantum case and discover that the choice of embedding maps is even more important in this case. We end in section \ref{disc} with a summary an outlook.

\section{The massless scalar field}\label{scalar}

As an instructive example consider the massless scalar field on a triangulation of flat Euclidean two--dimensional space. One\footnote{There are a number of discretizations for the 2D Laplacian, but there is not a naturally preferred one \cite{nofreelunch}.} discretization for the scalar field action on a triangulation is given by 
\ba\label{1}
S=\sum_\Delta  S_\Delta\;=\; \sum_\Delta \sum_{e\subset \Delta} \cot(\alpha_{\bar{e}}) (\phi_{s(e)} -\phi_{t(e)} )^2 \q ,
\ea
where $\alpha_{\bar e}$ is the angle at the vertex opposite the edge $e$, and $s(e),t(e)$ denote the source and target vertex of $e$ respectively. Note that in two dimensions the action for the massless scalar field is scale invariant.

The action (\ref{1}) is given as a sum over actions associated to the triangles. This additivity of the action will be essential for what follows. As we will see the action associated to one triangle can be interpreted as Hamilton's principal function for this triangle.

\begin{figure}[bt]
\begin{center}
    \psfrag{f1}{$\phi_1$}
    \psfrag{f2}{$\phi_2$}
    \psfrag{f3}{$\phi_3$}
    \psfrag{f4}{$\phi_4$}
     \psfrag{f0}{$\phi_0$}
    \psfrag{f2}{$\phi_2$}
    \psfrag{f3}{$\phi_3$}
    \psfrag{f0}{$\phi_0$}
       \includegraphics[scale=0.5]{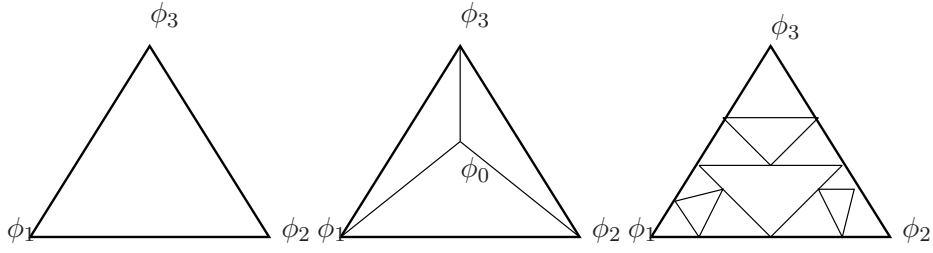}
    \end{center}
    \caption{\small \label{triang} Subdivisions of a triangle.}
\end{figure}

The dynamics defined by the action (\ref{1}) has the remarkable property to be invariant under the so-called 1--3 Pachner move. This move divides one triangle into three by introducing an inner vertex, see figure \ref{triang}. 
 (This has to be done such that the deficit angle at the inner vertex is zero, i.e. after the subdivision the outer triangle is still flat.)  Hamilton's principal function $H$, the action for the complex of three triangles, evaluated on a solution with given boundary data,  will then agree with the action for one triangle (which we identify with Hamilton's principal function for this triangle)
\ba\label{3}
{H_3}(\phi_1,\phi_2,\phi_3)\;=\; H_1(\phi_1,\phi_2,\phi_3) \q .
 \ea
Moreover the left hand side does not depend on where we place the inner vertex, that is the result is invariant under vertex displacement, which can be identified with a discrete notion of diffeomorphism symmetry \cite{rocek,dittrichreview,bahrdittrich1}.

The reason for this invariance is the simple solution for the field on the inner vertex given by
\ba\label{4}
\phi_0(x_1,x_2)=x_1 \phi_1 + x_2 \phi_2 + x_3(x_1,x_2) \phi_3
\ea
where the $x_i$ are barycentric coordinates of the triangle, i.e give the position of the inner vertex as a weighted average of the outer vertex positions, with $x_1+x_2+x_3=1$, so $x_3(x_1,x_2)=1-x_1-x_2$. This solution is a linear function in the coordinates and hence is also a solution of the continuum Laplace operator. For such a linear solution we can exactly give three boundary data which define the derivatives in the two coordinate directions. 

We can define this solution (\ref{4}) not only for the coordinate values of the actual inner vertex -- we can also take this as a continuum solution. 
This is the reason why Hamilton's principal function will be also invariant under further subdivisions of the triangle, including subdivisions of the boundary, as long as the boundary data  are chosen to be `edge wise' linear, i.e. 
\ba\label{5a}
\phi_k=x_l \phi_l + x_m \phi_m \q .
\ea
(Note that the set of coordinates with $x_i=0$ describes the line through the edge opposite the vertex $i$.) The boundary data we impose are the ones following from the dynamics, as these data are induced on the boundary by the solution (\ref{4}).

We conclude from this discussion that edge wise or piece wise linear boundary fields seem to be dynamically preferred. 

A two--dimensional triangulation can be changed into any other topologically equivalent triangulations by just applying $3-1$, $1-3$ and $2-2$ Pachner moves. Hence to  obtain full triangulation independence for our discrete theory we would just need invariance under the 2--2 move, which changes the way how two triangles are glued into a quadrangle. However the action (\ref{1}) is in general not invariant under the $2-2$ move (as the actions before and after the move display couplings along the two different diagonals). Additionally demanding invariance under the $2-2$ Pachner move and insisting on simple building blocks will lead to a non--local\footnote{The couplings will decay exponentially with the distance \cite{bietenholz}.}  action \cite{bietenholz,song}.



We will therefore employ a different strategy, allowing for more and more complicated building blocks. The discussion is easier on a regular lattice, so we start with actions on squares (which for simplicity we take to be also geometrical squares, i.e. equilateral). The justification for this choice is that our aim is to give a construction principle for discrete theories, such that the dynamics described does not depend on which choice of discretization or lattice one starts with, so we can choose a simple, regular lattice.

Hence we consider a lattice built out of squares. The action for one square can be obtained from the action for a triangular subdivision (\ref{1}). For a regular lattice we have (the lattice constant drops out as the action is scale invariant)
\ba\label{5}
S=\sum_{\square} S_\square \;=\; \sum_\square \sum_{i=1,\ldots 4}  (\phi_i-\phi_{i+1})^2   \q .
\ea
where $i=1,\ldots,4$ is a cyclic ordering of the vertices associated to the square $\square$, so addition of indices is to be understood mod $4$. The coupling along the diagonal drops out as $\cot(\alpha)=0$ for $\alpha$ orthogonal.

\begin{figure}[bt]
\begin{center}
    \psfrag{f1}{$\!\!\phi_1$}
    \psfrag{f2}{$\!\!\phi_2$}
    \psfrag{f3}{$\phi_3$}
    \psfrag{f4}{$\!\!\phi_4$}
     \psfrag{f0}{$\phi_0$}
    \psfrag{f12}{$\phi_{12}$}
    \psfrag{f23}{$\phi_{23}$}
    \psfrag{f34}{$\phi_{34}$}
    \psfrag{f41}{$~\!\!\!\phi_{41}$}
    \psfrag{g12}{$\gamma_{12}$}
 \psfrag{g23}{$\gamma_{23}$}
    \psfrag{g34}{$\gamma_{34}$}
    \psfrag{g41}{$~\!\!\!\!\gamma_{41}$}    
        \psfrag{g112}{$\gamma_{112}$}    
       \includegraphics[scale=0.5]{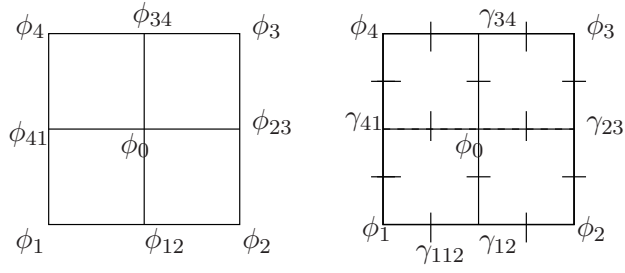}
    \end{center}
    \caption{\small\label{square} The boundary fields on a square. To find the fixed point action with eight boundary fields $\{\phi_i,\gamma_{ij}\}$ one would have to set i.e. $\gamma_{112}=0$ as it encodes the deviation from a piecewise linear field.}
\end{figure}

Now lets subdivide the square into four smaller squares. We obtain eight boundary fields and the field on the inner vertex $\phi_0$, with notation explained in figure \ref{square}. 
The eight boundary fields are given by the four fields on the corners of the composite square, which we denote by $\phi_i,i=1,\ldots 4$, and the boundary fields on the inner vertices of the subdivided edges which we denoted by $\phi_{ij}$ for the vertices between the corners $i$ and $j$.  

We solve the equation of motion for the inner vertex and insert this solution into the action for the four squares, which yields Hamilton's function depending on the eight boundary fields. Now we have to compare this to the original action, which is just a function of the four boundary fields.  Here we use the crucial insight from the previous discussion, which showed that piecewise linear fields define a special class of boundary data. So we define a new action as the Hamilton function, which we just computed, evaluated on piecewise linear boundary fields, that is we set $\phi_{ij}=\tfrac{1}{2}(\phi_i+\phi_j)$ for the fields on the midpoints of the edges. The resulting action
\ba\label{7}
S'_\square &=&(\phi_1-\phi_2)^2 +(\phi_2-\phi_3)^2+(\phi_3-\phi_4)^2 +(\phi_4-\phi_1)^2 -\alpha(\phi_1-\phi_2+\phi_3-\phi_4)^2 \nn\\
&=& S_\square -\alpha(\phi_1-\phi_2+\phi_3-\phi_4)^2 \q ,
\ea
where $\alpha=\tfrac{1}{2}$ for this first step, just differs by a total square from the initial action, introducing in particular  couplings along the diagonal.

This process can be repeated inducing a flow which only affects the value of $\alpha$. The fixed point $\alpha^*=\tfrac{2}{3}$ can be readily found.
%
This fixed point action enjoys an invariance under subdivisions of a certain type under the condition that the additional boundary data that result are fixed as (specific linear) functions of the initial boundary data. Later we will discuss in which sense this action already encodes the continuum dynamics for very special boundary data.

Moreover this fixed point is unique in the following sense. Parameterizing all quadratic (normalized) actions for a massless\footnote{which we define by an action which evaluated on constant fields gives zero} scalar field on a square by
\ba\label{8a}
S_\square=\sum_{i=1,\ldots,4} \left( \phi_i^2 +  a \phi_i \phi_{i+1} - (1+a) \phi_i \phi_{i+2}\right)
\ea
one will find that $a=-\tfrac{1}{2}$ is the only fixed point. This fixed point action agrees with the previous fixed point modulo a global rescaling.

Next we are going to construct an action which has the same invariance holding for more complicated boundary data, i.e. where the boundary fields can have a 'kink' at the midpoint of the edges. To each midpoint $(i,i+1)$ between the corners $i$ and $i+1$ we introduce the `fluctuation field'  $\gamma_{i,i+1}=\phi_{ij}-\tfrac{1}{2}(\phi_i+\phi_{i+1})$. 
As initial action we take Hamilton's principal function for the complex contracted out of four squares, defined by taking the fixed point action (\ref{9}) as the action for one square. So we just have to repeat the previous computation keeping the eight boundary fields general. 

 The resulting action is given by
\ba\label{9}
S_{\square^4} &=&S^*_\square  + \sum_{1\leq i\leq 4}  \left( \phi_i \gamma_{i,i+1} + \phi_i \gamma_{i,i-1}  -\phi_i\gamma_{i+1,i+2}  -\phi_i\gamma_{i-1,i-2} \right) \;+\;\nn\\
&&  \sum_{1\leq i\leq 4} \left( a_1\gamma_{i,i+1}^2+ a_2 \gamma_{i,i+1}\gamma_{i+1,i+2}+a_2 \gamma_{i,i+1}\gamma_{i-1,i}+\tfrac{1}{2}a_3 \gamma_{i,i+1}\gamma_{i+2,i+3} \right) \q 
\ea
with coefficients $a_\beta,\beta=1,2,3$, which flow under the coarse graining procedure: we glue four of such squares, solve for the fields on the (now five) inner vertices, and determine Hamilton's principal function on the subset of boundary data, where eight of the sixteen boundary fields are fixed as suitable averages of the eight remaining fields, see figure \ref{square}. We take Hamilton's principal function evaluated on such data as the new action for the square. 


The fixed point values for the coefficients $a_\beta,\beta=1,2,3$ can be given as the roots of a higher order polynomial, the numerical values are given by
\ba\label{10}
a_1: \, 2.5833 \rightarrow 2.2756 \, ,\q a_2:\, -1.5\rightarrow -1.2003 \,,\q a_3: -0.1666 \rightarrow -0.3427
\ea
where we gave the values for the initial Hamilton's principal function and the action/ Hamilton principal function at the fixed point.

One will notice that the zeroth order part -- the part quadratic in the $\phi$ fields -- does not change (if one starts with the previously computed fixed point action), nor does the part coupling the $\phi$ and the $\gamma$ fields. We want to emphasize that this is a special property of the  dynamics and choice of coarse boundary data considered here. To appreciate this point assume that we would treat the action (\ref{9})  in a perturbative expansion with small inhomogenities $\gamma$. To this end we introduce an expansion parameter $\epsilon$ so that the action has the form
\ba\label{10a}
S= \tfrac{1}{2} \phi \cdot M \cdot \phi +  \epsilon  \,\phi \cdot  N \cdot \gamma+ \tfrac{1}{2} \epsilon^2 \,\gamma \cdot K \cdot \gamma
\ea
where $M,N,K$ denote the quadratic forms specifying the dynamics.  For the solutions to the equation of motions we make the ansatz 
\ba\label{10b}
\phi={}^0\!\phi+\epsilon \,{}^1\!\phi \q , \q\q
\gamma={}^0\!\gamma
\ea
which have then to satisfy the equations of motion
\ba
 0=M\cdot {}^0\!\phi  + \epsilon \, \left( M\cdot {}^1\phi + N \cdot \,{}^0\!\gamma \right)  \label{10c}\q ,\q\q
 0=  \epsilon  \, {}^0\!\phi\cdot N + \epsilon^2\, \left(  {}^1\!\phi\cdot N  +  K \cdot {}^0\gamma\right) \q .
\ea
We can see that a perturbative solution is only possible if $ {}^0\!\phi\cdot N =0$ for solutions of the equations $M\cdot {}^0\!\phi=0$. This is a consistency conditions involving both $M$ -- that is the definition of the dynamics at zeroth order -- and $N$ -- that is the coupling between zeroth order and first order variables.
Similar consistency conditions occur for the discretization of systems with gauge symmetries in situations where these are broken by discretization, see the discussion in \cite{zako}.  

If the consistency conditions are satisfied the zeroth order solution will not be affected by the higher order terms (i.e. there are no backreaction effects). Remarkably these conditions are satisfied for the action (\ref{9}) and explain why only the quadratic part in $\gamma$ changes. We will however see that the first order term will change (slightly) if we include the higher order fluctuations.

We can introduce further fluctuation fields $\kappa_{i,i,i+1}$ and $\kappa_{i,i+1,i+1}$ which are associated to vertices subdividing the edges between $(i),(i,i+1)$ and between $(i,i+1),(i+1)$ respectively. As mentioned previously the first order part, which is quadratic in the $\gamma$-fields now does change, indicating that the $\gamma$ and $\kappa$ variables are not decoupled. The full fixed point action is quite lengthy, therefore we will
 just give the shifted values for the fixed point values of the coefficients for the $\gamma\cdot\gamma$ part
\ba\label{11b}
a_1:  2.2756\rightarrow2.2011 \, ,\q a_2:-1.2003\rightarrow-1.1805 \,,\q a_3: -0.3427\rightarrow-0.3562
\ea

Note that in the action (\ref{9}) 
all fields are coupled to each other, that is the theory is non--local  if one considers couplings between fields associated to one building block. On the other hand we still built the full dynamics out of local actions and building blocks which are glued to each other locally, namely by identifying fields on shared boundaries between building blocks. There are no couplings between fields on building blocks which do not share boundary fields. Such non--local couplings would however arise in a perfect action approach \cite{hasenfratz,bietenholz,song}, where one keeps the building blocks elementary.  This shift of non--locality can be understood as an example of the general method to avoid non-local couplings by introducing additional variables.

It should now be clear how to iterate this procedure. In each iteration we will allow more and more general boundary data. In the continuum limit we  can hope to obtain Hamilton's principal function for the boundary of a square.

\section{Scalar field with potential}\label{pot}

This method is also suitable for non--linear theories, e.g.  a massless scalar field with potential $\lambda \phi^4$. Here even the discretization on a triangle does not enjoy any more the invariance under subdivisions. Also there are different possibilities to discretize the potential term, for example one could take the average of the potentials of the fields or the potential evaluated on the average of the fields associated to one building block. We will see that with our method we can find a preferred discretization.

For the action on the square we make the ansatz that the potential term is a general homogenous polynomial of fourth degree in the fields. The kinetic term is given by the fixed point action for the massless field (\ref{7}):
\ba\label{12}
S_{\square,\lambda}&=&\!\!S_\square^* + \nn\\ &&\,\lambda a^2 \sum_i \big(      
 b_1 \phi_i^4 + b_2 \phi_i^3 \phi_{i+1} + b_2 \phi_i^3\phi_{i-1}+b_3 \phi_i^2\phi_{i+1}^2 +b_4 \phi_i^3\phi_{i+2}
 + b_5 \phi_i^2\phi_{i+1}\phi_{i+2} + \nn\\&&\,b_5 \phi_i^2\phi_{i-1}\phi_{i-2} + b_6 \phi_{i-1}\phi_i^2\phi_{i+1}+\tfrac{1}{2}b_7\phi_i^2 \phi_{i+2}^2 +\tfrac{1}{4} b_8 \phi_i\phi_{i+1}\phi_{i+2}\phi_{i+3}
            \big)\q\q
\ea
where $a$ is the lattice constant, so that the potential term is weighted with the area of the square. We proceed in the same way as in the previous section, that is we determine Hamilton's principal function on the subdivided square evaluated on special boundary data, which also here we take to be linear along the edges of the square. The only point one has to take care of is to replace $a\rightarrow a/2$ for the action of the smaller squares in the composite square in each coarse graining step. The full non-linear solution for the field on the inner vertex is a non-polynomial function of the boundary fields and in $\lambda$. To simplify the calculation we can expand in $\lambda$ and consider only the first order in $\lambda$, see section \ref{larger} for the second order result. The first order calculation implements also a truncation of the flow to polynomial actions up to degree four. As the quadratic part is already invariant only the coefficients $b_\beta$ will be changed. The solution to the fixed point equations is given by
\ba\label{13}
b_1=c,\q b_2=c, \q b_3=c,\q b_4=\tfrac{1}{4}c,\q b_5=\tfrac{1}{2}c,\q b_6=\tfrac{3}{4} c,\q b_7=\tfrac{1}{6}c,\q b_8=\tfrac{2}{3} c
\ea
where $c$ is a multiplicative constant that just rescales $\lambda$. (For the initial action (\ref{12}) with $b_1= 1,\,b_{i\neq1}=0$ we flow to $c=0.16$.) 

We see that we can define in this way a preferred discretization of the potential term for a given lattice. In the next section we will discuss whether this discretization encodes already the continuum result for edge--wise linear boundary data.


\section{Larger coarse graining steps}\label{larger}

In this section we are going to discuss the dependence of the sequence of fixed point action on changing the coarse graining scheme, in particular by taking larger coarse graining steps. This is actually very much related to the issue we discussed in section \ref{scalar}, namely whether the lower order results  changes if we take higher order fluctuations (more complicated boundary data) into account. 

We consider the scalar field (with $\lambda \phi^4$ potential) on a square with initial action (\ref{5}) and subdivide not only into four but into  sixteen squares. Again we put special boundary data. For the simplest case, we choose boundary fields which are linear along the edges of the bigger square. For the inclusion of the fluctuations $\gamma$ we allow a kink -- described by the $\gamma$--fields -- at the middle of the edges.

In general the fixed point actions will change as compared to the case, where we consider only the blocking of four squares into one square. The zeroth order result, i.e. the part which is quadratic in the fields $\phi_i$ does not change -- neither if we include fluctuations $\gamma$ or $\kappa$ into the boundary nor if we change the step size of the coarse graining procedure.


It is instructive to understand why this happens. The reason is in the choice of coarse boundary data -- as for the case of the triangle in section (\ref{scalar}) the piecewise linear boundary data are induced by a continuum solution which we can naturally obtain by extending the solution for the inner vertices and is given by
\ba\label{14}
\phi(x,y)= \sum_i (1-\,| x-x_i|)\,(1-\,| y-y_i|) \phi_i \q  .
\ea
Here $(x,y)$ are orthogonal coordinates on the larger square normalized such that the coordinate distance between neighbouring vertices of the larger square is equal to one. $(x_i,y_i)$ are coordinates of the vertex $i$. This solution holds for the actions  (\ref{7}) with any value of $\alpha$ and $\lambda=0$ and for the subdivision into any number of smaller squares. 
Indeed in the case that we have a larger square built out of more than four squares also the solution along the inner edges will be a linear function of some lengths parameter along this edge.



This reproduction of our choice of coarse boundary data along the inner boundaries is responsible for the stability of the zeroth order result under inclusions of higher order fluctuations. It also means that we actually found a continuum result: the fixed point action $S^\star_\square$ coincides with the continuum Hamilton's principal function evaluated on edge--wise linear boundary data.


The exact reproduction of our notion of coarse boundary data on the inner boundaries does not hold anymore if we include either the $\gamma$ fluctuations or the potential term $\lambda \phi^4$. For the latter note that the terms linear in $\lambda$, which we computed in section (\ref{pot}) will again remain stable both under the inclusion of $\gamma$ fields and under changing the blocking size from four to sixteen squares. The reason is that for the computation of Hamilton's function to first order in $\lambda$ one needs only the zeroth order (in $\lambda$) solution. Therefore the change will only start with the second order of $\lambda$. We therefore computed the fixed point action for the scalar field with $\lambda \phi^4$ potential to second order in $\lambda$, which is equivalent to a truncation to six order polynomials in the fields. We can compare the results with and without inclusion of $\gamma$ fluctuations into the boundary data, and with a blocking of four squares into one or with a blocking of sixteen squares into one. The final fixed point actions are very lengthy as for instance at second order all six degree polynomials in the eight boundary fields ($\phi_i$ and $\gamma_{ij}$) will appear. We will therefore just give the coefficients for the $(\gamma_{ij})^n,n=2,4,6$ terms in the action and the coefficient for the $\phi_i^6$ term for the various coarse graining schemes. For comparison in all cases the coefficient in front of $\phi_i^2$ is $\tfrac{4}{3}$ and the coefficient in front of $a^2\lambda\phi^4_i$ is $b_1=0.16$.

\begin{eqnarray}\label{tab1}
\begin{array}{|c||c|c|c|c|}
\hline
~&4\rightarrow 1;\, \phi_i& 4\rightarrow 1;\, \phi_i,\gamma_{ij}&16\rightarrow 1;\, \phi_i& 16\rightarrow 1;\, \phi_i,\gamma_{ij}\\ \hline\hline
\lambda^2a^4 \phi_i^6& -0.00389& -0.00498 &-0.00449&-0.00510\\ \hline
\gamma_{ij}^2&-&  2.27557&-&2.20273\\\hline
\lambda a^2\gamma_{ij}^4&-&0.06725&-&0.06012\\\hline
\lambda^2 a^4\gamma_{ij}^6&-&0.001020&-&0.001021\\
\hline
\end{array}
\end{eqnarray}

As we can see from the $\lambda^2\phi_i^6$ coefficients (and the other $\lambda^2$ coefficients in the $\phi$ fields not displayed here) the difference for the fixed point actions which include $\gamma$--boundary fields are smaller than for the cases without the $\gamma$--fields. Note also that the fixed point action computed from the coarse graining with $16\rightarrow 1$ square but without including $\gamma$ fields is different from the fixed point action computed in the $4\rightarrow 1$ scheme but with the inclusion of the $\gamma$ fields. (Compare also the $\gamma_{ij}^2$ coefficient computed with including the $\kappa$ fields, equation (\ref{11b}) and in the $16\rightarrow 1$ scheme.) 
The difference between the fixed point actions in the $4\rightarrow 1$ and in the $16\rightarrow 1$ scheme  involving the $\phi$ and $\gamma$ fields is however smaller compared to the case where only the $\phi$ fields are included. One would expect to obtain the continuum Hamilton's principal function if we consider larger and larger coarse graining steps. (This is just the requirement of having a discretization that gives the continuum result in the continuum limit.) In such a set--up we would however need to solve in one step for all the bulk fields at once.

Increasing the number of boundary fields instead of the blocking step size has the advantage that the computation proceeds step--wise. For example for the $16\rightarrow1$ scheme one has to solve for nine $(\phi)$ fields on the inner vertices, whereas for the $4\rightarrow1$ scheme and inclusion of $\gamma$ fields we just have to solve for five fields. Both schemes involve however the same number of boundary data (prior to restricting the boundary data to special `coarse' values). The reason why the second scheme involves less inner fields, is that we are using already Hamilton's function for the square with more boundary data, in which we already solved for the field on the inner vertex. This gives exactly the difference of the four fields, whose solutions we obtained in the previous computation.  In general we have a quadratic growth (in 2D) in the number of fields to solve for in the scheme with larger and larger blocking steps as compared to a linear growth in the scheme with including more and more boundary data.

To get an impression of the convergence rate of the lower order parts of the fixed point action under inclusion of more and more boundary data we consider a scalar field with mass in the $4\rightarrow 1$ blocking scheme and compute the fixed point action by taking only $\phi_i$ fields, $\phi_i$ and $\gamma_{ij}$ and finally $\phi_i,\gamma_{ij},\kappa_{iij}$ fields into account. Although we are just considering a quadratic action the full fixed point action will involve arbitrary high (even) powers of the mass. To make the computation feasible we truncate to $m^4$, that is to second order in the perturbation parameter $m^2$. (As for the $\lambda$ terms the $m^2 \phi_i^2$ terms of the action will remain stable under inclusion of higher order fields.) For comparison in all cases the coefficient in front of $\phi_i^2$ is $\tfrac{4}{3}$ and the coefficient in front of $a^2m^2\phi^4_i$ is $0.44444$.

\begin{eqnarray}\label{tab2}
\begin{array}{|c||c|c|c|c|}
\hline
~&4\rightarrow 1;\, \phi_i& 4\rightarrow 1;\, \phi_i,\gamma_{ij}&4\rightarrow 1;\, \phi_i,\gamma_{ij},\kappa_{iij}\\ \hline\hline
m^2a^2 \phi_i^2& 0.44444& 0.44444 &0.44444\\ \hline
m^4a^4 \phi_i^2& -0.01799& -0.02126 &-0.02172\\ \hline
m^2a^2\gamma_{ij}^2&-& 0.21639&0.20764\\\hline
m^4 a^4\gamma_{ij}^2&-&-0.00913&-0.00929\\\hline
m^2 a^2\kappa_{iij}^2&-&-&0.06700\\\hline
m^4 a^4\kappa_{iij}^2&-&-&-0.00195\\\hline
\hline
\end{array}
\end{eqnarray}

As expected the change in the coefficient of the $m^4a^4\phi_i^2$ term is decreasing, from an order of $10^{-3}$ to an order of $10^{-4}$ between the different truncations. Thus we can hope on a convergence of the coefficients to some (continuum) value.

In conclusion we have to expect `back reaction' contributions from the inclusion of more inhomogeneous boundary data onto the terms in the fixed point action involving more homogeneous fields. This applies even for the linear dynamics in the case that the corresponding field modes do not completely decouple, as is the case between the $\gamma$ and $\kappa$ fields. Such back reaction effects could be avoided by changing to decoupled modes, i.e. for a translation invariant boundary the Fourier modes. In this case the higher (order) modes will not appear at all in the solution for the bulk fields -- as long as these modes are set to zero in the boundary data. Hence the boundary data reproduce itself on the inner boundaries, as discussed above. 

Of course the construction of decoupled modes will only be possible for free theories. For interacting theories we might be able to identify modes, that is a decomposition of boundary fields, which only couple weakly with each other. The (coarse) boundary data will only be approximately reproduced on inner boundaries, but we can expect a fast convergence of the terms involving only lower order modes under inclusion of higher (order) modes into the boundary data.
 
 The advantage of the scheme proposed here is that one can determine the size of the corrections at least from one order to the next order, so that one can judge the reliability of the approximation scheme.

\section{Other coarse graining schemes}\label{other}

Here we will discuss  a coarse graining scheme with a different choice of coarser boundary data. This choice follows more closely the usual understanding of coarse graining, e.g.  fields in a region are averaged to a new field, constant on the region averaged over. 

\begin{figure}[bt]
\begin{center}
    \psfrag{f1}{$\!\!\phi_1$}
    \psfrag{f2}{$\!\!\phi_2$}
    \psfrag{f3}{$\!\!\phi_3$}
    \psfrag{f4}{$\!\!\phi_4$}
     \psfrag{f0}{$\phi_0$}
    \psfrag{f12}{$\phi_{12}$}
    \psfrag{f23}{$\phi_{23}$}
    \psfrag{f34}{$\phi_{34}$}
    \psfrag{f41}{$~\!\!\!\!\!\phi_{41}$}
    \psfrag{f11}{$\!\!\phi_{11}$}
 \psfrag{f21}{$\phi_{21}$}
    \psfrag{f22}{$\phi_{22}$}
     \psfrag{f31}{$\phi_{31}$}
    \psfrag{f32}{$\phi_{32}$}
    \psfrag{f42}{$~\!\!\!\!\!\phi_{42}$}    
        \psfrag{f5}{$\!\!\!\!\phi_{5}$}    
          \psfrag{f6}{$\!\!\phi_{6}$}    
            \psfrag{f7}{$\!\!\phi_{7}$}    
              \psfrag{f8}{$\!\!\phi_{8}$}    
       \includegraphics[scale=0.5]{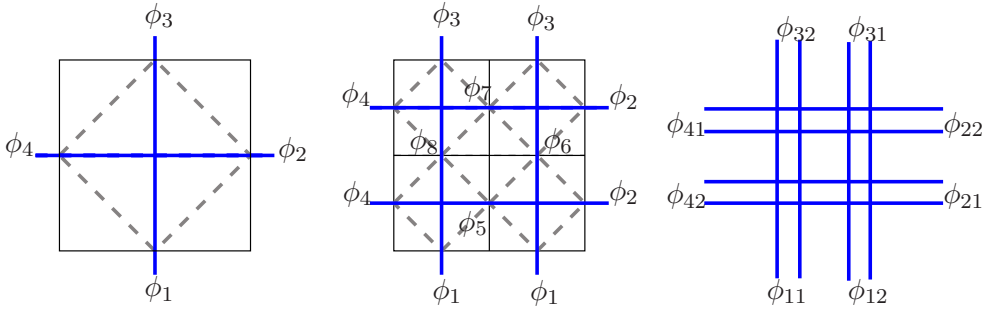}
    \end{center}
    \caption{\small \label{sqtnw}The tensor network scheme for the coarse graining of a scalar field. The original square is rotated by $45^\circ$, so that now the boundary fields can be associated to the edges of an unrotated square. The gluing of these new squares corresponds to combining tensors of rank four. The corresponding tensor network is indicated by blue lines. Bulk fields $\phi_5,\ldots,\phi_8$ are now associated to the inner edges of this tensor network.  
    Coarser boundary data correspond to piece wise constant fields. In the figure on the right we depicted the set--up for coarse graining with eight boundary fields -- fields on pairs of outer legs are set to be equal.}
\end{figure}

Hence we replace  for the coarser boundary data `piecewise linear'  fields with `piecewise constant' fields. For this we have also to change the geometrical set-up, as the fields will be now associated to the (midpoints of the) edges. Such a set--up corresponds more closely to a tensor network renormalization approach, see \cite{levin,wen,eckert} and figure \ref{sqtnw}.

 Coarser boundary data are now obtained by setting fields on a subdivided boundary edge to be equal. 
  There is one subtlety one has to take care of: to find a fixed point one has to introduce a scale $s$ between the new average boundary fields and the to be averaged finer boundary fields, i.e. $\phi_{is}=s (\phi_{i1}+\phi_{i2})$ for the case that the boundary fields agree on pairs of boundary edges $(i1,i2)$. To find the fixed point action, one has to normalize one of the coefficients in the action after each renormalization step to one. At the fixed point one can then determine the scaling $s$  by finding the normalization constant. This scale has to be properly taken into account if a potential term is added: as we can interpret $s$ as a length scale, fields carry the dimensions of inverse lengths. Hence if we have to multiply the terms $\phi^2$ with $s^{-2}$,  we should multiply  terms $a^4\phi^2$ with $s^2$ whereas terms $a^2 \phi^2$ are dimensionless and will not be rescaled.
  
  We determined the fixed point action for the scalar field with mass, in a scheme where four squares are blocked into one square. We computed the fixed point action for three different set of boundary data: starting with four boundary fields $\{\phi_{i},i=1,\ldots 4\}$ which we interpret as edge--wise constant fields, to a set $\{\phi_{i1},\phi_{i2},i=1,\ldots,4\}$ of eight boundary fields which we can interpret as piecewise constant fields along the edges, and finally a set $\{\phi_{i1},\phi_{i2},\phi_{i3},\phi_{i4}\}$ of sixteen boundary fields. We will just display the coefficients for these three cases after restricting the boundary values to the simplest case, i.e. setting $\phi_{i1}=\phi_{i2}=\ldots$ in the fixed point action. The resulting actions have to be normalized again in the way described in the previous paragraph, so that the actions can be compared with each other. The initial action is given by
  \ba\label{f1}
  S_{\square}=\sum_i (\phi_i^2 -\phi_i\phi_{i+1}+a^2m^2\phi_i^2) \q .
  \ea
   In the following table we will display the results for the three different sets of boundary fields with the notation
 \ba\label{f2}
  S_{\square}=\sum_i \big(\phi_i^2  + b_1 \phi_i\phi_{i+1} +\tfrac{1}{2}b_2 \phi_i\phi_{i+2} &+&a^2m^2 ( c_1 \phi_i^2  + c_2\phi_i\phi_{i+1} + \tfrac{1}{2}c_3 \phi_i\phi_{i+2}  )\nn\\
  &+&a^4m^4 ( d_1 \phi_i^2  + d_2\phi_i\phi_{i+1} + \tfrac{1}{2}d_3 \phi_i\phi_{i+2}  )
  \big) \q .
 \ea
 Because of the rescaling procedure the coefficient of the term $\phi_i^2$ will be always one, furthermore we will have $b_2=-2(1+b_1)$ which ensures that the massless part vanishes, if evaluated on a constant boundary field.
  
\begin{eqnarray}\label{tab3}
\begin{array}{|c||c|c|c|c|}
\hline
~&4 \,\text{bdry fields}&8 \,\text{bdry fields}&16 \,\text{bdry fields}\\ \hline\hline
b_1& 0.928& 0.945 &0.958\\ \hline
c_1& 0.493& 0.534 &0.536\\ \hline
c_2&0.408& 0.370&0.366\\\hline
c_3&0.197&0.192&0.191\\\hline
d_1&-0.189&-0.203&-0.261\\\hline
d_2&-0.260&-0.283&-0.373\\\hline
d_3&-0.181&-0.214&-0.286\\\hline
\hline
\end{array}
\end{eqnarray}

 From table  (\ref{tab3}) we can see that we do not have stability of the coefficients at zeroth order, as was the case for the scheme with piecewise linear boundary fields. We explained in section \ref{larger} that this cannot be expected as piecewise constant fields do not appear as solutions on the inner edges. The coefficients for the kinetic part of the action and the ones in front of the $m^2$ terms do converge, although slower than in the scheme with piecewise linear boundary fields. Moreover the coefficients in front of the $m^4$ terms show an unstable behaviour. In summary the scheme with piecewise linear fields seems to be better suited for the dynamics of a scalar field (perturbed around the massless case).

In general we have to choose carefully a notion of coarse boundary data, which will underly the coarse graining scheme. This notion should be adjusted to the dynamics -- ideally one should capture the relevant degrees of freedom.

Another possibility to determine a notion of coarse boundary data is to follow the tensor network renormalization algorithm, where the truncation to coarse boundary data is entirely determined by the dynamics.  This procedure was however developed not for classical theories but for quantum ones (with discrete configurations, like Ising spin systems), in which the partition function is rewritten as a contraction of a tensor network. Contracting just four tensors $T$ to a new tensor $T'$, this tensor will have the double number of indices, which correspond to the double number of boundary data in each coarse graining step in our case. Thus one has to truncate the tensor down to its original size as otherwise the number of indices grows exponentially with the number of iterations. This truncation is motivated by finding the most relevant modes in the tensor $T$ that contribute mostly to the partition function. The general problem in identifying these modes is that there is no canonical definition of eigenvalue decomposition for tensors, as there is for matrices. Thus a procedure has been designed that introduces auxiliary matrices that allow for an eigenvalue decomposition \cite{levin,wen}.  In general, the identification of the `most relevant' modes leads to field redefinitions, that may also depend on the  renormalization step. The  modes, identified as relevant, are included in the next iteration step, whereas the irrelevant ones are discarded. 

As the entire procedure is not so much motivated by finding the boundary wave function (see however the discussion in \cite{levin}) for given boundary data, one usually does not keep track of these field redefinitions. This could however be done in principle. The information gained in this way could be used to identify a notion of coarse boundary data that is adjusted to the dynamics and therefore might also change during the iteration procedure. The resulting notion will of course depend on the truncation scheme for the tensors. On the other hand if one has reason to believe that a given notion of coarsening captures the relevant degrees of freedom, this notion can be used to define a truncation scheme for the tensor network algorithm. Section \ref{quantum} will discuss a possibility to extend our classical scheme to quantum mechanics, and we will see that identifying a proper notion of coarse boundary data becomes even more essential.

\section{Cylindrical consistency}\label{cyl}

In the discussion so far a choice of coarse boundary data was essential. We can formalize this notion using the concept of cylindrical consistency. This is a key idea in 
loop quantum gravity \cite{al,thomasbook}, which allows the construction of a Hilbert space encoding the continuum configurations through an inductive limit of a family of Hilbert spaces that are associated to (finite) graphs. There is a natural (partial) ordering of graphs into coarser or finer graphs. Cylindrical consistency implements the requirement that a computation (of expectation values, transition amplitudes etc.) for a given state should not depend on whether this computation is performed in a Hilbert space associated to a coarser or finer graph, as long as both graphs are fine enough to describe the given state. 


Following this idea we introduce a (partial order) directed set structure on the set ${\cal B}$ of boundaries $b$, where $b\prec b'$, if $b'$ is a refinement of $b$.  To every boundary we associate a configuration space ${\cal C}_b$, which typically is a direct product of some basic configuration space ${\cal C}$.

The notion of coarse boundary data will be encoded in the choice of a family of embedding maps. For every pair $b\prec b'$ we define an injection
\ba\label{01b}
\iota_{bb'}:{\cal C}_{b}\rightarrow {\cal C}_{b'}  \q .
\ea
where the image of $\iota_{bb'}$ determines the `coarse boundary data' in the configuration space ${\cal C}_{b'}$. This family of injections needs to be consistent in the sense that for $b\prec b' \prec b''$ we should have
\ba\label{02b}
\iota_{bb''}=\iota_{b'b''}\circ \iota_{bb'} \q .
\ea
Such a consistent family allows the construction of an inductive limit
\ba\label{02bb}
{\cal C}_{\text{ind}}=\cup_b {\cal C}_b/\sim \q ,
\ea
which consists of equivalence classes of elements in a disjoint union of configuration spaces ${\cal C}_b$ over all boundaries $b\in {\cal B}$. Two elements $c_b$ and $c_{b'}$ are equivalent if there exist a $b''$ with $b\prec b''$ and $b'\prec b''$ and $\iota_{bb''}(c_b)=\iota_{b'b''}(c_{b'})$ holds. In other words two configurations defined on two different boundaries are equivalent if there is a finer boundary onto which the two configurations can be embedded and if these embeddings happen to agree.


A family of functions  $\{F_b\}_{b\in {\cal B}}$ is cylindrically consistent  on the inductive family defined by $({\cal C}_b,\iota_b)$, if
\ba\label{03a}
F_b=\iota^*_{bb'}F_{b'} \;,\q  \text{i.e.} \q\q  F_b(c)=F_{b'}( \iota_{bb'}(c)) \q\q \forall c\in {\cal C}_b
\ea
where $\iota^*_{bb'}$ is the pullback of $\iota_{bb'}$.  This just implements that the value of the function $F$ should not depend on the representative on which one chooses to evaluate. 

The procedure proposed in this work attempts to construct Hamilton's principal function as a cylindrically consistent object: Assume that we do have the continuum Hamilton's function at our disposal. In the language we are using here, the continuum is given by the refinement limit of the boundaries $b$, that is on a heuristic level we can imagine the continuum as an element $c$ for which $b \prec c$ for all finite boundaries in ${\cal B}$. Now we can pull--back  the continuum Hamilton's function to any coarser boundary and the family of function so obtained will be cylindrically consistent. To be more precise the continuum limit of a function is actually defined through the condition of having a cylindrically consistent family of functions.

    We compute this continuum limit  iteratively starting from an action $S_b$ (identified with Hamilton's principal function) on some given boundary $b$. This is used to compute Hamilton's principal function on some finer boundary $b'$, we will denote the result by $S_{b'}^b$.  For the fixed point action $S^*_b$ we demand that
\ba\label{03b}
\iota^*_{bb'}S_{b'}^b=S^*_b \q .
\ea
If the fixed point action $S^*_b$ in (\ref{03b}) does not depend on the choice of finer boundary $b'$, then it is the $b$--component of the cylindrically consistent family of Hamilton's principal functions we are looking for: in the limit that $b'$ approaches the continuum we converge to the continuum Hamilton's principal function and the pull back to the coarser boundary $b$ defines the $b$--component of the corresponding cylindrically consistent family of functions.

We have seen that in the case of the massless free scalar field and for the zeroth order coefficient such an independence holds, but that in general one can only hope for approximate (or a convergence) of results. Here the choice of the finer boundary $b'$ determines the order of the approximation for the $b$--component of the cylindrical consistent family of Hamilton's functions. 

We should remark that (\ref{03b}) is to be understood for systems without an explicit scale, such as the massless scalar field. Note that gravity, where the metric is a dynamical variable, is in this sense also without a scale (if one does not expand on a background)  \cite{ditt}. This also holds for parametrized systems, where embedding coordinates are added as dynamical variables  \cite{ditt}. In all these cases a notion of scale can be replaced by the notion of coarser and finer boundary data.

In the case of scale dependent actions, which we denote by $S_{b,a}$, for example the scalar field with potential, we have to replace (\ref{03b}) by  
\ba\label
{04b}
\iota^*_{bb'}S_{b',La}^{b,a}=S^*_{b,La}
\ea
where the lattice constant in $b'$ is $La$ and $S_{b',L a}^{b,a}$ is Hamilton's principal function computed from the action $S_{b,a}$. Even more generally the embeddings $\iota^*_{bb'}$ could also depend on the two scales involved.

In the previous scalar field examples, based on piecewise linear fields, the basic configuration space is associated to a vertex and given by $\Rl$. The configuration space for a boundary with $N$ vertices is given by $\Rl^N$. We considered the set of boundaries of a square with a given number of repeated subdivisions of the edges into two halves. $b'$ is a refinement of $b$ if $b'$ can be obtained from $b$ by such a subdivision. Let us consider the case that $b'$ is obtained from $b$ by one subdivision. We label the vertices in $b$ by indices $i$ (cyclically ordered) and the additional vertices in $b'$, which are between $i$ and $i+1$ (mod the number of vertices in $b$)  by $(i,i+1)$. The injection map is then given by defining the fields on the additional vertices as average of the neighbouring vertices, i.e.
\ba\label{02}
\iota_{bb'}(\phi_1,\phi_2,\phi_3,\ldots)=(\phi_1,\tfrac{1}{2}(\phi_1+\phi_2),\phi_2,\tfrac{1}{2}(\phi_2+\phi_3),\phi_3,\ldots)  \q .
\ea
Iterating this procedure one obtains a directed set of boundaries and can define the inductive limit of configuration spaces.

\section{Remarks on the quantum theory case}\label{quantum}

As mentioned in the previous section the notion of cylindrical consistency is crucial in the loop quantization of gravity. We therefore remark that the notion of cylindrical consistency for Hamilton's principal function translates into the notion of a cylindrical consistent (path integral) measure in the quantum theory, a notion proposed in \cite{benloops}. Such a cylindrical consistent measure would allow to make the path integral well defined \cite{benloops}. Here we will only sketch the idea and point out one of the main issues one would have to address even in the case of free theories. 

 Hamilton's principal function associated to a region with a boundary is the classical analogue  to the path integral over this region.  The result of this path integral as a function of the boundary data is the (generalized) boundary wave function \cite{oeckl}. 
 Note that -- given a prescription for the path integral this boundary wave function is the unique {\it physical} state associated to this region.\footnote{Usually one would think of the path integral to propagate states from one boundary to a second one. Here we have only one boundary and the wave function corresponds to the Hartle--Hawking `no boundary' wave function. This is fully consistent with the classical phase space one would associate to such a boundary, as such a classical phase space is totally constrained \cite{philipp2}.)}

For the following it is useful if we take a dual point on the path integral and rather see it as an (anti--linear) functional on the (kinematical) boundary Hilbert space.
That is we  assume that we can associate to every boundary of a region a kinematical boundary Hilbert space ${\cal H}_{bound}$, that encodes all states that can in principle occur kinematically. A well defined quantum theory should then associate an amplitude map to every region, that is it defines a map on every boundary Hilbert space:
\ba\label{q1}
{\cal P}_{bound}: {\cal H}_{bound} \mapsto \Cl \q .
\ea
As we map to $\Cl$ the co--kernel of this map is one--dimensional, defining the physical wave function, which we mentioned before.
 The maps ${\cal P}$ for unions of regions have to satisfy consistency conditions which are generalizations of the usual composition rule for the path integral \cite{oeckl}.  Classically these consistency conditions correspond to the additivity of the action (or the composition rule for Hamilton's principal function).

These considerations also hold for the case that we replace space time with a lattice or more general discretization. In this case the boundaries are discrete boundaries $b$ to which we associate Hilbert spaces ${\cal H}_b$. As for the classical case we have to choose injection maps $\iota_{bb'}$ that  embed the Hilbert space ${\cal H}_b$ into ${\cal H}_{bb'}$. Such injections can be naturally constructed if we have projections $\pi_{b'b}:{\cal C}_{b'}\rightarrow {\cal C}_b$ on the configuration spaces at our disposal: assuming a polarization in which the quantum states are functions on the configuration space we can use the pullback of the projections to define injections:
\ba\label{q2}
\iota_{bb'}:&&{\cal H}_b\rightarrow {\cal H}_{b'} \nn\\
&& \psi_{b}\mapsto \psi_{b'} \q \text{where} \q\q\psi_{b'}(c_{b'})=\psi_b(\pi_{b'b}(c_{b'})) \q .
\ea
The reader will note that in the classical case we also used injections instead of projections. In the case we discussed here we can however obtain projections by `forgetting' the fluctuation variables. I.e. in the case that coarse graining was based on piecewise linear fields and for a refinement which doubled the number of fields, $\{\phi_i\}$ to $\{\phi_i,\phi_{i,i+1}\}$, we perform a variable transformation $\gamma_{ij}=\phi_{ij}-\tfrac{1}{2}(\phi_i+\phi_j)$ and define the projection as
\ba\label{q3}
\pi_{b'b}(\phi_1,\gamma_{12},\phi_2,\ldots)=(\phi_1,\phi_2,\ldots) \q .
\ea
Such a projection satisfies $\pi_{b'b} \circ \iota^{class}_{bb'}=\text{Id}_{b}$ where here the $\iota^{class}_b$ is the classical embedding map for the configuration spaces. 

Such inductions which are based on projections are also used for the construction of the Hilbert space in Loop Quantum Gravity \cite{al}, see also \cite{carlos}.

  A measure (which here will be just understood as linear functional) on a family of inductive Hilbert spaces can be defined by providing a representation on each of the Hilbert spaces ${\cal H}_b$. Such a family of measures $\{\mu\}_b$ is cylindrically consistent if
  \ba\label{q4}
\mu_b(\psi_b)=\mu_{b'}(\iota_{bb'}(\psi_b)) \q .
\ea
The Ashtekar--Lewandowki measure \cite{al, thomasbook} for the kinematical Hilbert spaces in loop quantum gravity satisfies such consistency relations. 

One way to define the dynamics (and the physical inner product) would be via the path integrals acting as  functionals 
on these kinematical Hilbert spaces\footnote{technically on some dense subspace of these Hilbert spaces} as in (\ref{q1}). Cylindrical consistent dynamics would then mean to have a cylindrical consistent family of (anti--linear) maps
\ba\label{q5}
{\cal P}_b:{\cal H}_b\rightarrow \Cl \q .
\ea

As an example we can again consider the massless free scalar field. The kinematical Hilbert space associated to the boundary with $N$ vertices will be $L^2(\Rl^N)$, here we will have the case $N=4$ and $N=8$. We define the path integral map for the boundary $b$ of the basic square as\footnote{The complex conjugation of the kinematical wave function can be interpreted as the kinematical wave function being associated with the other orientation of the boundary \cite{oeckl}. }
\ba\label{q6}
{\cal P}_b(\psi_b)&=&\frac{1}{N_b}\int \prod_{i=1,\ldots,4} \bd \phi_i  \,\, e^{iS_b(\phi_i)} \, \overline{\psi_b}(\phi_i) \q 
\ea
where $N_b$ is a normalization factor.
The path integral map for a refinement $b'$ that subdivides a  square into four would be given by 
\ba\label{q7}
{\cal P}_{b'}(\psi_{b'})&=&\frac{1}{N_{b'}}\int  \bd\phi_0\prod_{i=1,\ldots, 4} \bd\phi_i  \bd \gamma_{i,i+1}   \,\, e^{iS_{b'}(\phi_i, \gamma_{i,i+1},\phi_0)} \, \overline{\psi_{b'}}(\phi_i,\gamma_{i,i+1}) 
\ea
where $\phi_0$ is the scalar field on the inner vertex. $\gamma_{i,i+1}$ are the variables introduced in (\ref{q3}) and $S_{b'}$ is here the action for four squares obtained by summing the basic action $S_b$ over the four squares. If we define the injections $\iota_{bb'}$ as pullback of the projections (\ref{q3}) we obtain for the pullback of the path integral map from $b'$ to $b$
\ba\label{q8}
(\iota_{bb'})^*{\cal P}_{b'} \,(\psi_b)={\cal P}_{b'}(\iota_{bb'}(\psi_b))=\frac{1}{N_{b'}}\int  \bd\phi_0\prod_{i=1,\ldots, 4} \bd\phi_i  \bd \gamma_{i,i+1}   \,\, e^{iS_{b'}(\phi_i, \gamma_{i,i+1},\phi_0)} \, \overline{\psi_{b}}(\phi_i)  \q .
\ea 
For cylindrical consistency of the path integral maps defined in this way we would need 
\ba\label{q8b}
\frac{1}{N_b}\int \prod_{i=1,\ldots,4} \bd \phi_i  \,\, e^{iS_b(\phi_i)} \overline{\psi_b}(\phi_i) =\frac{1}{N_{b'}}\int  \bd\phi_0\prod_{i=1,\ldots, 4} \bd\phi_i  \bd \gamma_{i,i+1}   \,\, e^{iS_{b'}(\phi_i, \gamma_{i,i+1},\phi_0)} \,{\overline \psi_{b}}(\phi_i) 
\ea
for a suitable class of test states $\psi_b$. As we are considering free field theory the integral over the $\gamma$ fields and $\phi_0$ is a (analytically continued) Gaussian integration. Indeed, as pre--factors arising from these integrations can be absorbed into the normalizations $N_b,N_{b'}$ we can just consider the extremization of $S_{b'}$ with respect to $\phi_0$ and then with respect to the fields $\gamma_{i,i+1}$. 

Compared to the classical coarse graining procedure we therefore do not set the $\gamma_{i,i+1}=0$ but have to find the extrema of Hamilton's principal function $H(\phi_i,\gamma_{ij})=S_{b'}(\phi_i,\gamma_{ij},\phi_0^{\text{sol}})$ with respect to these variables $\gamma$. We then obtain a new action, by evaluating Hamilton's function on these extrema. This new action will depend just on the four fields $\phi_i$ and we can repeat the procedure and in this way obtain a renormalization flow of actions. 

The resulting flow does however not converge to a fixed point. Rather a rescaling fixed point is reached, in which the action is rescaled by a given number (smaller than one). Indeed compared to the classical case we do not fix the boundary values of $\gamma$, but do minimize further the action with respect to these variables. The values of the boundary fields $\gamma$ obtained in this way are however not reproduced, if we put the resulting squares again together, so that we have now a variation of the fields $\gamma$ in the bulk. This is different from the classical case, where the boundary values $\gamma = 0$ are reproduced in a larger square. Indeed if we include in this procedure higher order fields the (rescaling) fixed point, which differs from the classical one, will change towards the location of the classical fixed point. 

Although one could deal with such a rescaling fixed point, the issue seems rather that we do not reproduce the classical result in the semi--classical limit. The key point responsible is the choice of embedding map, which also describes the choice of boundary conditions (for the vertices carrying $\gamma$ fields) imposed on the quantum mechanical wave function, which we are trying to find. This point constituted also the main motivation for the introduction of the density matrix renormalization group procedure \cite{white}:  find the ground state of a given Hamiltonian in a given block with some fixed boundary conditions (for instance that the wave function vanishes on the boundary) . Putting the blocks together and finding the ground state again for the larger block it might happen that the wave function evaluated on the now inner boundaries differs very much from the values prescribed by the conditions on the smaller blocks. Even for the free particle in one dimension one will find a ground state wave function that vanishes on the ends of the smaller intervals, versus a maximum of the ground state wave function assumed on the midpoint of the doubled interval. This issue is solved by a kind of bootstrap approach -- also the appropriate boundary conditions have to be constructed by considering not only the block itself but the block inside an environment (built from the same block). 

For our case (of a free field) we can attempt to reproduce the classical coarse graining by choosing an appropriate embedding map. In this case we have to pre--guess the correct vacuum for the finer degrees of freedom. We can choose the embedding as
\ba\label{q9}
\psi_b(\phi_i)\mapsto \psi_b(\phi_i)\exp(i T(\gamma\cdot\phi, \gamma\cdot \gamma)) \q .
\ea
 The condition (\ref{q8b}) turns into
 \ba\label{q10}
 \frac{1}{N_b}\int \prod_{i=1,\ldots,4} \bd \phi_i  \,\, e^{iS_b(\phi_i)} \overline{\psi_b}(\phi_i) =\frac{1}{N_{b'}}\int  \bd\phi_0\!\!\!\!\prod_{i=1,\ldots, 4}\!\!\! \bd\phi_i  \bd \gamma_{i,i+1}   \,\, e^{iS_{b'}(\phi_i, \gamma_{i,i+1},\phi_0)-iT(\gamma\cdot\phi, \gamma\cdot \gamma)   } \,{\overline \psi_{b}}(\phi_i) \,. \nn\\
 \ea
We can fix $T$ such that the term $T(\gamma\cdot\phi,\gamma\cdot\gamma)$ exactly cancels the $\gamma$ dependence in the action $S_{b'}$. If $S_b$ is the (classical) fixed point action, we reproduce also cylindrical consistency in the quantum case.
(This will lead to a divergence in the integral over $\gamma$. In the case of loop quantum gravity the corresponding integration is over a compact space. Also here we could (Bohr) compactify $\Rl$, equivalent with a regularization of the integral to a finite interval $[-C,C]$ and a normalization by $1/2C$, letting then $C\rightarrow\infty$.) 

We can interpret the choice of embedding (\ref{q9}) as selecting the correct vacuum for the finer degrees of freedom $\gamma$. Only test states $\psi_b \cdot e^{iT}$ which have already the correct semi-classical behaviour with respect to the $\gamma$ degrees of freedom, are propagated in the same way on the finer boundary, as the states $\psi_b$ on the coarser boundary.

In conclusion the choice of embedding seems to be even more essential in the quantum case. For the construction of these embedding maps methods from the density matrix renormalization group or tensor network based renormalization \cite{levin,wen,vidal}, which can be seen as a descendent of  the density matrix renormalization group procedure, could be helpful. We would not have encountered problems with (\ref{q8b}) , if the finer and coarser degrees of freedom where completely decoupled. Achieving such a decoupling is one of the motivations behind the approach of entanglement renormalization \cite{vidal}. The tensor networks can in this case be understood as parametrizing an ansatz for a physical wave function.

Here we started with an embedding map inspired from loop quantum gravity, which we argued is not an appropriate one for coarse graining the free scalar field (on a flat background). This does not mean that the loop quantum gravity embedding map might not be useful for coarse graining -- as we argued the embedding should depend on the system and the dynamics it describes. Indeed there are uniqueness results available for the loop quantum gravity embedding, and the vacuum underlying this embedding \cite{flost}, which are based on the requirement of a (spatially diffeomorphism symmetry covariant) irreducible representation of the kinematical observable algebra and a (spatially) diffeomorphism invariant vacuum. On the other hand, it is not guaranteed that this embedding will allow for the construction of a cylindrically consistent measure encoding the dynamics via a path integral, in the way we outlined here, i.e. via a coarse graining procedure. As we have seen in the scalar field example we might need already some information on the physical vacuum for the finer degrees of freedom.  The question is whether we can encode this physical vacuum into an inductive limit structure for the physical\footnote{In gravity physical states have to be annihilated by the diffeomorphism and Hamiltonian constraints} Hilbert space. The crucial question will be whether such a structure exists and how it relates to the inductive limit structure of the kinematical Hilbert space. 
Note also that alternatives for the kinematical Hilbert space have been explored \cite{tim,carlos}.

\section{Summary, open issues and outlook}\label{disc}

We proposed a coarse graining scheme, in which the central object is Hamilton's principal function for the classical case and the physical wave function for the quantum case. Both can be taken to define an improved or perfect discretization \cite{hasenfratz,bahrdittrich2,seb1}: Hamilton's principal function can be used as the action associated to a building block with the corresponding boundary and the physical wave function  can be used as the path integral amplitude for this building block. 

The main properties for this coarse graining scheme are:
\begin{itemize}\parskip -2pt
\item Following the tensor network renormalization \cite{levin}, instead of introducing non--local couplings between building blocks we introduce building blocks with more and more boundary data. The non--localities are shifted to occur in-between the boundary data associated to one building block. An advantage of this scheme is that using these building blocks to define a discretization the amount of non--locality (i.e. the maximal lattice distance at which fields are coupled) can be chosen. This choice also determines the truncation of the renormalization flow, as couplings over larger lattice distances are neglected. 
\item
The quality of this truncation can be tested by considering the higher order corrections. The approximations to Hamilton's function for a given boundary $b$ are specified by finer boundaries $b'$, see formula (\ref{03b}). Such an approximation can be understood as an effective action which takes the corrections into account, that arise from the additional degrees of freedom in $b'$ as compared to $b$. Taking the refinement limit of $b'$ one obtains the continuum limit of Hamilton's function for boundary data described by $b$.

\item Crucial for the coarse graining scheme is the choice of coarse boundary data encoded into a family of embedding maps. These embedding maps provide also an interpretation of discrete boundary data as configurations in a continuum theory. The embedding maps have to be adjusted to the dynamics, ideally the notion of coarse boundary data should be reproduced by the dynamics of the system. Equivalently the choice of coarse boundary data should correspond to an ordering of the variables into finer and coarser degrees of freedom which are dynamically only weakly coupled.
\item The scheme proposed here seems to be especially suited for systems without an explicit (lattice) scale, which includes gravity systems \cite{ditt}. Here a notion of scale can be replaced by a notion of coarser and finer  boundary data. A certain truncation might give reliable result for boundary data up to a given complexity.
\end{itemize}

There are many open issues left to be understood, in particular for the quantum theory case. 
The main point is how to select best the embedding maps. Here methods developed in condensed matter and quantum information theory \cite{white,vidal} will be helpful. One question to address will be in which situations it is appropriate to determine the truncation scheme (or embedding maps) completely by the dynamics, as is done in \cite{levin,wen,eckert,vidal}  or whether some predefined embedding might be sufficient. For the investigation of phase transition one would rather expect to need truncations that are determined by the dynamics. The second possibility might arise if we aim at determining a perfect discretization of a system in a well defined phase. This will be investigated for simplified models \cite{finite,eckert}  in \cite{toappear}.

For instance the Ashtekar--Lewandowski measure (and the embedding maps and kinematical vacuum on which this measure is based), used in Loop Quantum Gravity, corresponds to the high temperature fixed point of lattice gauge theory \cite{benloops}. Here an interesting question is whether we could construct a Hilbert space and measure corresponding to the low temperature fixed point of lattice gauge theory, which is actually given by a topological theory, namely $BF$--theory, see for instance \cite{finite}. $BF$ theory underlies spin foam quantization, hence it would be natural to consider a coarse graining procedure with corresponding embedding maps. In this case the ('kinematical') vacuum state should correspond to a physical state of $BF$ theory, which in (2+1) D even coincides with the physical state of gravity. As shown in \cite{karim} the construction of this physical state can be obtained starting from the usual loop quantum gravity Hilbert space. Hence we see that it is actually possible to obtain a physical vacuum, starting from a kinematical Hilbert space, which is based on embeddings which rather encode a kinematical vacuum describing a very different phase. The question is however, which are the appropriate embedding maps to choose for the coarse graining procedure proposed here. For other LQG related Hilbert spaces, based on 'condensate' vacua, describing non--degenerate geometries, see \cite{tim}.

Another important question to address is whether one can estimate and obtain a bound for the `back reaction' effects,  i.e. the change of the lower order results, obtained for very coarse boundary data, if finer and finer boundary data are included. This would already entail the existence of such a continuum limit, which in itself is a highly non--trivial question, in particular if it comes to the quantum theory. The advantage of the scheme here is that it provides a systematic way to improve the truncation in steps, and that the higher order corrections can be checked explicitly.

Let us mention some possible applications of this scheme. 

The main motivation is the construction of perfect discretizations for (quantum) gravity models. 
Perfect discretizations have been proposed for gravity \cite{alg,bahrdittrich2} in order to avoid the breaking of diffeomorphism symmetry for generic discretizations \cite{bahrdittrich1}. The notion of diffeomorphism symmetry for discretizations \cite{rocek,bahrdittrich1} is a very powerful one as it implies discretization (or triangulation) independence \cite{seb1,seb2,ditt,zako}. 
 In the scheme proposed here the requirement of discretization or triangulation independence is imposed by computing fixed point actions at every step. We can then hope to obtain triangulation independence up to a given measure of fineness of the boundary data. In its simplest form this is already realized for (classical) Regge gravity \cite{regge}, namely for boundary data that lead to flat \cite{jimmy,seb2} or constantly curved solutions \cite{newregge} (for gravity with a cosmological constant). Correspondingly a natural choice of an embedding map would be to implement piecewise flat or piecewise constantly curved geometries as coarser boundary configurations, see also \cite{zip}. 

More generally the methods proposed here could be used to define improved discretizations for partial differential equations, if these can be derived from an action principle. The use of renormalization methods to obtain improved discretizations has been proposed for instance in \cite{goldenfeld,degenhard}, in particular in \cite{degenhard} there are also embedding (and projection) maps used. The difference with the scheme here is the utilization of the action principle, the emphasis on interpreting discrete boundary data as elements in finer configuration spaces, as well as the possibility to improve the discretization in steps and to control the amount of non--local couplings.  

A particular problem for discretizations, related to the breaking of diffeomorphism symmetry is the loss of energy momentum conservation, see for instance \cite{marsden,zako}. Perfect discretizations would avoid such a breaking as these are defined to display continuum physics. This problem can be studied in parametrized field theory, where diffeomorphism symmetry is added through the introduction of embedding variables \cite{ishamkuchar}. The problem of energy conservation translates then into the problem of broken diffeomorphisms and in the canonical set--up to an anomalous constraint algebra. With the methods proposed here we can hope to obtain a discretization that is perfect for a certain class of boundary data and for which anomalies in the constraint algebra can be avoided. On a general method to obtain a discrete canonical formalism with the same (diffeomorphism) symmetries as the covariant one, see \cite{bahrdittrich1,philipp1,philipp2}, in particular \cite{philipp2} is applicable to boundaries with varying number of variables, which are also occurring here.

Of course we hope that the scheme proposed here can also be applied to quantum gravity models, in particular spin foams \cite{carlobook,perezreview}. Although spin foams, being seen as a quantum version of Regge calculus have been originally defined on simplicial lattices only, recently models on arbitrary lattices (two--complexes) have been constructed \cite{warsaw,frankandco}. These also provide an example of `more complicated' (non--simplicial) building blocks, which allow the allocation of finer and finer boundary data. We propose here that the amplitudes for such building blocks can be constructed through a coarse graining process as outlined in this work. See also \cite{eckert} for a formulation of spin foams as tensor networks and a coarse graining procedure for simplified models. There are also formulations available which define spin foams as generalized lattice gauge theories \cite{pfeiffer,claudio,benloops,frankandco}, which again can be extended to any two--complex. These formulations might allow to systematically investigate which notion of coarse boundary data, or embedding maps, will lead to interesting results. In particular \cite{finite,eckert,frankandco} provide a range of simplified models for which this question can be more easily explored. 

The standard LQG embedding provides a well defined choice and is strongly motivated by (spatial) diffeomorphism symmetry. As outlined in \cite{zip}, at least on the classical  level  there are different interpretations possible\footnote{These different interpretations arise trough an interpretation of the discrete phase space as a reduction of a continuum phase space by a certain set of gauge symmetries. Different gauge fixings lead to different interpretations.} of how the discrete data are to be understood in a continuum phase space. 
We suggest here that an embedding of discrete data into a continuum configuration or Hilbert space should be determined by the dynamics. Such an embedding would be rather associated to a physical vacuum and a physical Hilbert space -- as we indicated in section \ref{quantum} it would encode information on the vacuum wave function for the finer degrees of freedom.

Finally, we would like to mention that the notion of coarse and finer boundary degrees of freedom can be brought together with symmetry reduction and applications in (quantum) cosmology. Very coarse states would correspond to homogeneous configurations, finer states would describe inhomogenities. Indeed in the context of loop quantum cosmology \cite{martin} there are different suggestions how the kinematics and dynamics can be obtained from the full theory  \cite{tim,cosmo,frankh} and how to include corrections from inhomogenities, with \cite{frankh} discussing (cylindrical) consistency conditions for the truncation of a given theory to homogeneous degrees of freedom.

\section*{Acknowledgements}

 I am very much indebted to Benjamin Bahr for many discussions, comments on the draft and suggesting the notion of cylindrical consistency for the coarse graining scheme. I furthermore thank Jan Ambj\o rn, Laurent Freidel, Frank Hellmann, Carlo Rovelli, Lee Smolin and Sebastian Steinhaus for discussions.
Research at Perimeter Institute is supported by the Government of Canada through Industry Canada and by the Province of Ontario through the Ministry of Research and Innovation.

\vspace{1cm}

\appendix


\end{document}